\documentstyle[prb,aps,graphicx]{revtex}

\def\be{\begin{equation}}
\def\ee{\end{equation}}
\def\ba{\begin{eqnarray}}
\def\ea{\end{eqnarray}}

\begin{document}
\draft
\title{Floating Wigner molecules and possible phase transitions in quantum dots}
\author{S. A. Mikhailov\footnote{Present address: Max-Planck Institut f\"ur Festk\"orperforschung, Heisenbergstr. 1, 70569 Stuttgart, Germany} and K. Ziegler}
\address{Theoretical Physics II, Institute for Physics, University of Augsburg, 86135 Augsburg, Germany}
\date{\today}
\maketitle

\begin{abstract}
A floating Wigner crystal differs from the standard one by a spatial averaging over positions of the Wigner-crystal lattice. It has the same internal structure as the fixed crystal, but contrary to it, takes into account rotational and/or translational symmetry of the underlying jellium background. We study properties of a floating Wigner molecule in few-electron spin-polarized quantum dots, and show that the floating solid has the lower energy than the standard Wigner crystal with fixed lattice points. We also argue that internal rotational symmetry of individual dots can be broken in arrays of quantum dots, due to 
degenerate ground states and inter-dot Coulomb coupling. 
\end{abstract}


At sufficiently low densities a system of interacting electrons on a uniform positive background forms a correlated Wigner-crystal state. The idea of an electron crystal was originally proposed by Wigner,\cite{Wigner34} and then extensively studied in three-\cite{Fuchs35,Wigner38,Clark58,Brout59,Coldwell60,Carr61a,Carr61b,deWette64,Horn67,Edwards68,Bagchi69,Foldy71,Aguilera85} and two-dimensional\cite{Chaplik72,Meissner76,Bonsall77,Ceperley78,Fukuyama79,Maki83,Widom88,Tanatar89,Chui95,Yi98} (2D) electron systems (experimental realizations are electrons on liquid Helium \cite{Grimes79} and semiconductor \cite{Yoon99} surfaces). Recently, an interest aroused to the formation and properties of Wigner molecules in semiconductor quantum dots -- zero-dimensional systems with a finite number $N$ of 2D electrons.\cite{Bolton93,Bedanov94,Creffield99,Reimann00,Egger99a,Egger99b,Filinov01,Hausler00,Reusch01,Mikhailov02b}

A standard Wigner-crystal ground-state trial wave function $\Psi_{WC}({\bf r}_1,{\bf r}_2,\dots,{\bf r}_N)$ (for a fully-spin-polarized electron system) has the form of a Slater determinant constructed from single-particle orbitals $\psi({\bf r})$ centered at points, corresponding to equilibrium positions of classical particles, 

\be\Psi_{WC}(\{{\bf r}_i\},\{{\bf R}_j\})=\frac{1}{\sqrt{N!}}\det_N\left| \psi({\bf r}_i- {\bf R}_j) \right|.
\label{trialfunction0}
\ee 
In a circular parabolic quantum dot with a small number $N$ ($\le 8$) of 2D electrons the vectors ${\bf R}_j$ form symmetric polygonal configurations with $N_c$ electrons in the center,\cite{Bolton93,Bedanov94}
\ba
{\bf R}_j&=&R(\cos \theta_j ,\sin \theta_j),\ \ \theta_j=2\pi (j-1)/N, \ \ j=1,\dots,N,\ \ N\le 5, \nonumber \\
{\bf R}_1={\bf 0},\ \ {\bf R}_j&=&R(\cos \theta_j ,\sin \theta_j),\ \ \theta_j=2\pi (j-2)/(N-1), \ \ j=2,\dots,N,\ \ N=6,7,8,
\label{sites}
\ea
$N_c=0$ at $N\le 5$, and $N_c=1$ at $6\le N\le 8$ (for quantum dots with more particles see examples of classical configurations in \cite{Bolton93,Bedanov94}). In a macroscopic 2D electron system vectors ${\bf R}_j$ form a triangular lattice,\cite{Meissner76,Bonsall77}
\be
\{{\bf R}_j\} \sim a\left(l_1+l_2/2,\sqrt{3}l_2/2\right),{\ \ }n_sa^2=2/\sqrt{3},
\label{tr-lattice}
\ee
with the period $a$ corresponding to the average area density of 2D electrons $n_s$ ($l_1$ and $l_2$ are integer). 

The equilibrium classical configurations (\ref{sites}), (\ref{tr-lattice}) are usually considered to be fixed with respect to some fixed external reference frame. This leads, in fact, to a violation of the rotational (in circular quantum dots) and translational (in infinite 2D systems) symmetry of the ground state, and results in an angular- and position-dependent ground-state electron density. The trial many-body wave function (\ref{trialfunction0}) in the dots (in an infinite 2D electron system) is not the eigenfunction of the total angular momentum $\hat L_{tot}$ (of the total momentum $\hat {\bf P}_{tot}$), although the Hamiltonian has the rotational (translational) symmetry and commutes with the corresponding operators (here $\hat L_{tot}\equiv \hat L_z^{tot}$, where $z$-axis is perpendicular to the plane of the 2D system).

The symmetries can be restored in (\ref{trialfunction0}) by averaging over the 
corresponding positions of the vectors ${\bf R}_j$. Such a floating solid was discussed in the literature in connection with other many-body problems, see e.g. \cite{Feenberg74,Bishop82}. In this paper we discuss some general properties of a floating Wigner molecule in a circular parabolic quantum dot, and of a floating Wigner crystal in a macroscopic 2D electron system. We show that $\{{\bf R}_j\}$-averaged Wigner-crystal wave functions can be constructed as eigenfunctions of the total angular/total momentum operators, and that these eigenstates possess the symmetry of the Hamiltonian. We perform specific calculations for the energy of few-electron quantum dots, in the fixed and floating Wigner-crystal states, and show that the $\{{\bf R}_j\}$-averaging does lead to an essential reduction of the variational ground-state energy. For two-, three- and four-electron parabolic quantum dots we compare our variational results with exact results available in the literature. We also show that the internal rotational symmetry of the wave functions in circular dots can be broken if the ground state is degenerate with respect to the total angular momentum $L$: a superposition of eigenfunctions with different $L$ breaks the symmetry without increasing the energy of the state. We argue that this leads to phase transitions in quantum-dot arrays with spontaneous symmetry breaking.


Consider a circular $N$-electron parabolic quantum dot. Let ${\bf R}_j^\alpha=R_j(\cos (\theta_j+\alpha) ,\sin (\theta_j+\alpha))$ be a set of vectors rotated around the origin by an angle $\alpha$ with respect to (\ref{sites}). Consider a function
\be
\Psi_L(\{{\bf r}_i\})=\int_{-\pi}^\pi\frac{d\alpha}{2\pi}e^{iL\alpha}\Psi_{WC}(\{{\bf r}_i\},\{{\bf R}_j^\alpha\}),
\label{L-functions}
\ee
obtained from (\ref{trialfunction0}) by averaging over $\alpha$ with the weight function $e^{iL\alpha}$, and labeled by an arbitrary integer $L$. The functions (\ref{L-functions}) have the same internal structure as (\ref{trialfunction0}), so that electron-electron correlations are taken into account at the same level of accuracy, and satisfy the eigenvalue equation 
\be
\hat L_{tot}\Psi_L(\{{\bf r}_i\})=\hbar L \Psi_L(\{{\bf r}_i\}),
\label{Ltoteigeneq}
\ee
with $L$ being the total angular momentum quantum number. The fixed Wigner-crystal wave function (\ref{trialfunction0}) can be presented in the form of an infinite sum of eigenfunctions $\Psi_L$ with all possible values of $L$,

\be
\Psi_{WC}(\{{\bf r}_i\},\{{\bf R}_j\})=\sum_{L=-\infty}^\infty \Psi_L(\{{\bf r}_i\}).
\label{sumL}
\ee
These results obviously remain valid if the trial wave functions are multiplied by additional Jastrow factors (see e.g. \cite{Ceperley78,Tanatar89}) which depend only on relative coordinates of electrons.

For a quantum dot with $N\le 8$ electrons we calculate expectation values $E=\langle\Psi|\hat H|\Psi\rangle/\langle\Psi|\Psi\rangle$ of the Hamiltonian $\hat H$ in the states (\ref{L-functions}) and (\ref{trialfunction0}), and directly compare the energies $E_L$ and $E_{WC}$ of the floating and fixed Wigner-molecule states. The Hamiltonian of the dots has the form 

\be
\hat H =\sum_{i=1}^N \left (\frac {\hat {\bf p}_i^2}{2m^\star} + \frac {m^\star\omega_0^2 {\bf r}_i^2}{2} \right)+ \frac 12\sum_{i\neq j=1}^N \frac{e^2}{\kappa|{\bf r}_i-{\bf r}_j|} ,
\label{ham}
\ee
where $m^\star$ is the effective mass of 2D electrons, $\kappa$ is the dielectric constant of the host semiconductor, and $\omega_0$ characterizes the confinement. For the single-particle orbitals $\psi({\bf r})$ we take gaussians

\be
\psi({\bf r})=\frac 1{\sqrt{\pi}l}\exp\left(-\frac {{\bf r}^2}{2l^2}\right),
\ee
with the width $l$ being either equal to the oscillator length $l_0=\sqrt{\hbar/m^\star\omega_0}$, or considered as a variational parameter. For the vectors ${\bf R}_j$ we assume the classical equilibrium configurations (\ref{sites}).

Due to the symmetry of configurations (\ref{sites}) not all values of $L$ are allowed in Eq. (\ref{L-functions}). For instance, if $N=2$ and the system is fully spin-polarized (the total spin is $S=N/2=1$) the number $L$ can take only odd values; otherwise the integral (\ref{L-functions}) vanishes. For the ground state it is reasonable to assume the smallest possible value of the total angular momentum, so that we get $L=\pm 1$. As known from exact-diagonalization calculations,\cite{Merkt91} at $S=1$ the ground state of a two-electron quantum dot has indeed the total angular momentum $|L|=1$. Similarly, if $N=3$, allowed values of $L$ are $0,\pm 3,\pm 6,\dots$, and in the ground state of the fully polarized ($S=3/2$) three-electron dot we have $L=0$. This also agrees with exact-diagonalization calculations \cite{Mikhailov02b}. In general, for $N$-electron dots with $N$ from 2 to 8, the total spin $S=N/2$, and classical configurations (\ref{sites}), we get the following ground-state total angular momenta

\ba
L=&(N-N_c)/2, \ \ &{\rm if}\ (N-N_c)={\rm even}, \nonumber \\
L=&0, \ \ \ \ \ \ \ \ \ \ \ \ \ \ \ &{\rm if}\ (N-N_c)={\rm odd},
\label{allowedL}
\ea
where $N_c=0$, if $N\le 5$, and $N_c=1$, if $6\le N\le 8$. 

\begin{figure}
\includegraphics[width=8.2cm]{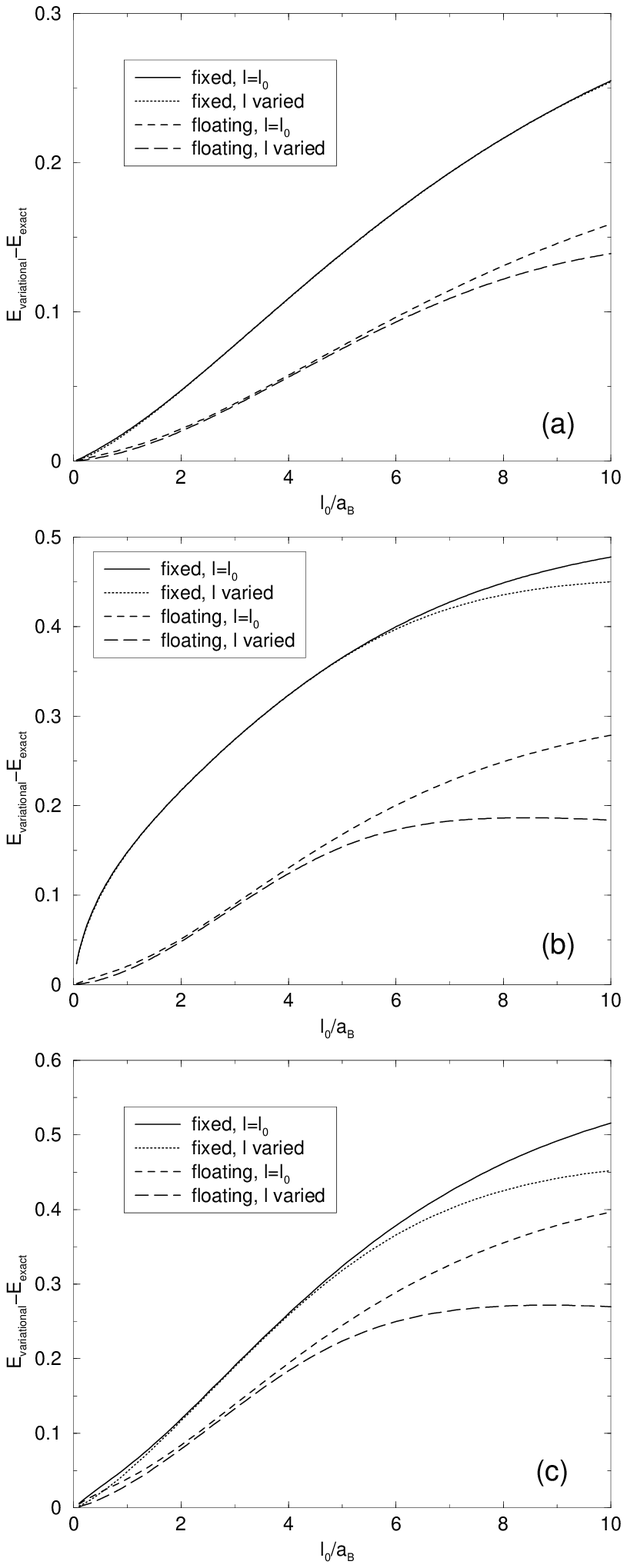}
\caption{Energy difference $E_{WC}-E_{exact}$ (curves labeled as ``fixed'') and $E_L-E_{exact}$ (curves labeled as ``floating'') for parabolic quantum dots with (a) $N=2$, (b) $N=3$ and (c) $N=4$ spin-polarized electrons, as a function of the Coulomb interaction parameter $l_0/a_B=\sqrt{e^2/a_B\hbar\omega_0}$ (energy unit is $\hbar\omega_0$, $a_B$ is the effective Bohr radius). For both trial wave functions the curves without and with optimization over the variational parameter $l$ are shown (the curves labeled as ``$l=l_0$'' and ``$l$ varied'' respectively). Exact-diagonalization results are taken from Ref. \protect\cite{Merkt91} ($N=2$), Ref. \protect\cite{Mikhailov02b} ($N=3$), and Ref. \protect\cite{MikhailovBe} ($N=4$). }
\label{23el}
\end{figure}

\begin{figure}
\includegraphics[width=8.2cm]{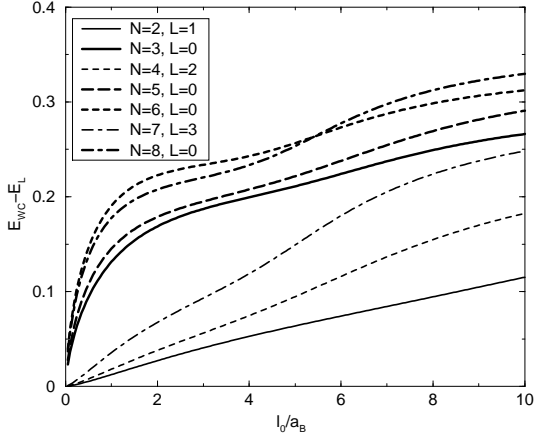}
\caption{Energy difference $E_{WC}-E_{L}$, in units $\hbar\omega_0$, as a function of the interaction parameter $l_0/a_B$, for parabolic quantum dots with $N=2$ to 8. For both trial wave functions optimization over the variational parameter $l$ have been performed.}
\label{28el}
\end{figure}

The energies $E_L$ (with $L$ from Eq. (\ref{allowedL})) and $E_{WC}$ have been calculated with the help of general formulas obtained in Ref. \cite{Mikhailov01}. Figure \ref{23el} shows the difference $(E_{WC}-E_{exact})$, $(E_L-E_{exact})$ between the energies of the fixed and floating variational Wigner-crystal states and exact-diagonalization results from Refs. \cite{Merkt91,Mikhailov02b,MikhailovBe}, for $N=2$, $3$, and $4$ spin-polarized electrons, $S=N/2$. Figure \ref{28el} exhibits the difference $(E_{WC}-E_L)$ between the energies of the fixed and floating Wigner-molecule states, for parabolic quantum dots with $N$ from 2 to 8 electrons. One sees that in all the cases the ${\bf R}_j$-averaging leads to an essential improvement (by about 0.2--0.3 $\hbar\omega_0$) of the variational ground state energy, and that in general the energy gain increases with the number of particles in the dots. Another (quite natural) finding is that for $\Psi_L$ states with zero total angular momentum (four upper thick curves in Figure \ref{28el}) the energy gain is substantially larger than for the states with $L\neq 0$ (three lower thin curves). The difference between the variational and exact ground-state energies is reduced by a factor of $\approx 2$ in the two-, three-, and four-electron quantum dots, Figure \ref{23el}, due to the floating version of the Wigner-molecule trial wave function. 

\begin{figure}
\begin{center}
\includegraphics[width=7.5cm]{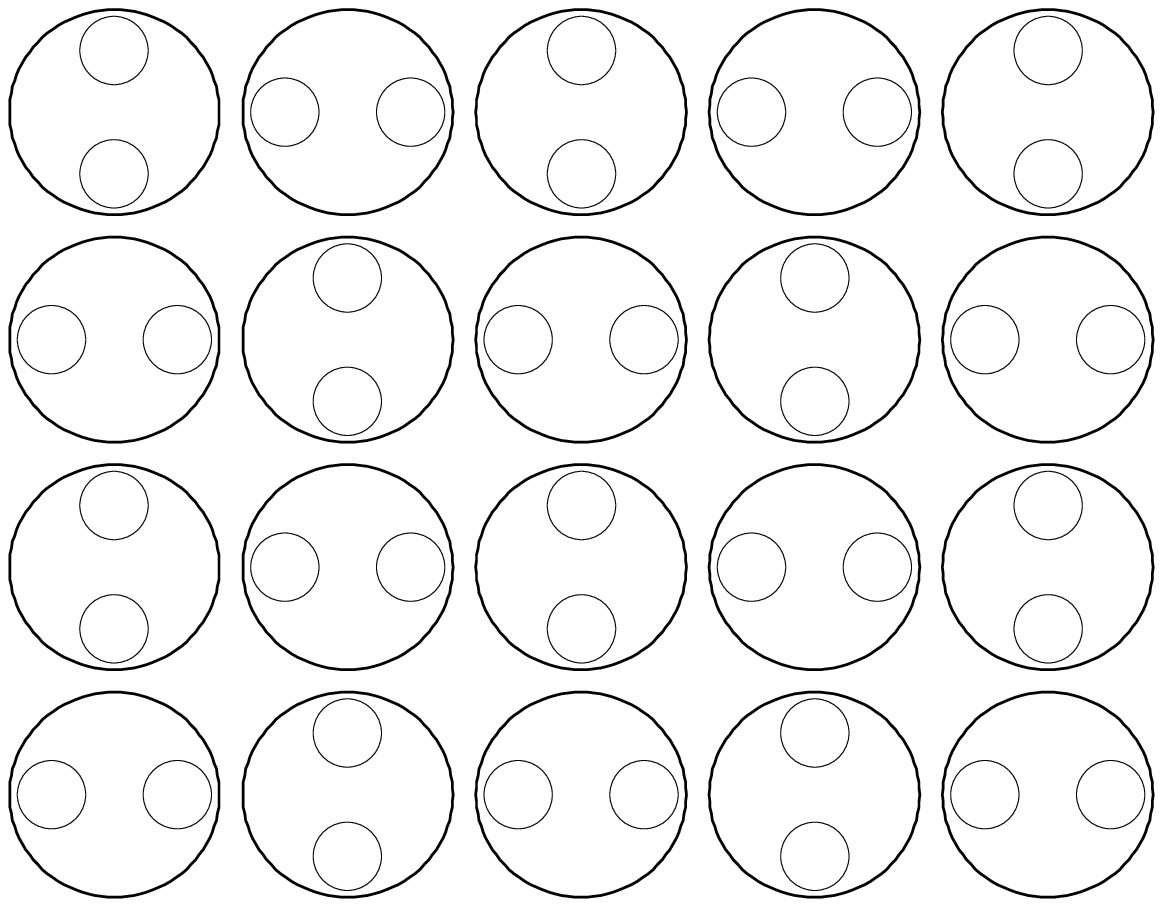}
\caption{Possible phase transition to a broken-symmetry state in an array of Coulomb-coupled quantum dots. It is assumed that there are two electrons in each dot (thin circles show the maxima of the electron density), and that electron spins are polarized.}
\label{dotpictures}
\end{center}
\end{figure}

The states $\Psi_L$ are eigenstates of the total angular momentum $\hat L_{tot}$, Eq. (\ref{Ltoteigeneq}), and the density of electrons $n_L({\bf r})$ in these states does not depend on the angular coordinate, $n_L(r,\theta)=n_L(r)$. However, if the ground state is degenerate, as it is the case, for instance, at $L\neq 0$ ($N=2$, 4, and 7), one can construct solutions
\be
\Psi_{mixed}=C_L\Psi_L+C_{-L}\Psi_{-L},
\label{mixedsolution}
\ee
which have the same (ground-state) energy $E_{mixed}=E_{\pm L}$, but an angular-dependent density $n_{mixed}(r,\theta)$. A possible mixed state is $(\Psi_L+\Psi_{-L})/\sqrt{2}$ with $\langle L\rangle=0$. Such broken-symmetry ground states can thus exist in a circularly symmetric quantum dot, depending on initial conditions. In a {\em single} dot the energies of symmetric and broken-symmetric solutions are the same, but in the presence of a small perturbation the symmetry-broken state may have the lower energy (an evident example is a circular dot with ``disorder'', e.g. with an asymmetrically located impurity). Consider for instance an {\em array} of circular dots with a weak inter-dot Coulomb coupling, and assume that each dot contains two spin-polarized electrons (no tunneling is assumed between the dots). Then the structure schematically shown in Figure \ref{dotpictures} will obviously have the lower energy than the state with angular independent density in each dot. A weak inter-dot Coulomb interaction in arrays of dots can thus lead to phase transitions with a spontaneously broken phase. 

Results presented in formulas (\ref{L-functions}) -- (\ref{sumL}) can be easily generalized to the case of other electron systems on the jellium background. For example, for an infinite 2D electron system (with periodic boundary conditions) the floating Wigner crystal wave function can be written as

\be
\Psi_{\bf K}(\{{\bf r}_i\})=
\int d{\bf a} e^{i{\bf K}\cdot{\bf a}}\Psi_{WC}(\{{\bf r}_i\},\{{\bf R}_j+{\bf a}\}),
\label{K-functions}
\ee
where the integral is taken over the (large) area of a sample. The functions (\ref{K-functions}) are the eigenfunctions of $\hat {\bf P}_{tot}$,
\be
\hat {\bf P}_{tot}\Psi_{\bf K}(\{{\bf r}_i\})=\hbar{\bf K}  \Psi_{\bf K}(\{{\bf r}_i\}),
\ee
with $\hbar {\bf K}$ being the total momentum quantum number. The fixed Wigner crystal wave function (\ref{trialfunction0}) is expanded in terms of $\Psi_{\bf K}$ as

\be
\Psi_{WC}(\{{\bf r}_i\},\{{\bf R}_j\})=\frac 1{(2\pi)^2}\int d{\bf K} \Psi_{\bf K}(\{{\bf r}_i\}).
\ee
The floating Wigner crystal wave functions (\ref{K-functions}) give the position-independent density of 2D electrons. However, if the ground state is degenerate with respect to ${\bf K}$, a superposition of such degenerate ground states can again lead to a modulated density.

To summarize, we have studied some general properties of a floating Wigner crystal in parabolic quantum dots and other electron systems. In few-electron dots the rotationally-invariant floating solid was proved to have a substantially lower energy than the fixed Wigner molecule given by the trial wave function of
Eq. (\ref{trialfunction0}). In arrays of quantum dots the inter-dot interaction may lead to a phase transition to the ground state with broken rotational symmetry in individual dots. 

The work was supported by the Deutsche Forschungsgemeinschaft through the SFB 484. We thank Karl-Heinz H\"ock for useful discussions.


\end{document}